\documentclass[journal,10pt]{IEEEtran}
\ifCLASSINFOpdf
\else
\fi
\usepackage{amsfonts,amsmath,amssymb}
\usepackage{mathtools}
\usepackage{mdwmath}
\usepackage{cuted}
\usepackage{mdwtab}
\usepackage{amsthm}           
\usepackage{bm}               
\usepackage{units}            
\usepackage{graphicx}
\usepackage{epstopdf}
\usepackage[linesnumbered,ruled,vlined]{algorithm2e}
\usepackage{url}

\usepackage{paralist}
\usepackage{color}
\usepackage{authblk}
\usepackage[caption = false]{subfig}
\usepackage{acronym} 

\newcommand{\Fig}[1]{Fig.~\ref{fig:#1}}


\begin{document}
%
\title{UAVs as Mobile Infrastructure: Addressing Battery Lifetime}
%
%
%

\author{Boris Galkin
        and~Luiz~A. DaSilva
}

\affil{CONNECT, Trinity College Dublin, Ireland \\
\textit{E-mail: \{galkinb,dasilval\}@tcd.ie}}

\maketitle

\begin{abstract}
Unmanned aerial vehicles (UAVs) are expected to play an important role in next generation cellular networks, acting as flying infrastructure which can serve ground users when regular infrastructure is overloaded or unavailable. As these devices are expected to operate wirelessly they will rely on an internal battery for their power supply, which will limit the amount of time they can operate over an area of interest before having to recharge. In this article, we outline three battery charging options that may be considered by a network operator and use simulations to demonstrate the performance impact of incorporating those options into a cellular network where UAV infrastructure provides wireless service.
\end{abstract}


%
\IEEEpeerreviewmaketitle

\section{Introduction}

In recent years remote-controlled flying devices (UAVs) have expanded from the domain of military applications into civilian markets \cite{Fitzpatrick_2018}, with millions of consumer-grade UAVs being sold every year around the world. This evolution is attributed to a number of recent technological developments which have made it possible to develop small, affordable UAVs capable of carrying out a variety of tasks using on-board devices. These tasks include the use of UAVs in emergency applications, for industrial and agricultural inspections, and for package delivery. There is a growing interest among the wireless research community in the possibility of using UAVs as flying infrastructure acting alongside, or in place of, terrestrial networks, in a variety of scenarios \cite{Zeng_20162}.

UAV-mounted communications infrastructure is a complete paradigm shift which can bring several key benefits over the existing mobile network infrastructure, including:
\begin{enumerate}
\item UAVs, due to their airborne nature, can establish much higher quality channels to a terrestrial receiver, with significantly lower signal attenuation. Whereas a macro base station mounted on a building rooftop will experience non-Line-of-Sight (LOS) propagation conditions to its associated ground users due to the buildings in the way, the UAV can adjust its height to hover high above such obstacles. This benefit is particularly significant in urban areas with high building density.
\item The UAVs, as they can move on command, can optimise their locations in real-time with respect  to the location of the traffic demand, and can then readjust as the demand changes. This is in stark contrast to existing infrastructure, which is fixed in place and relies on careful site planning on the part of the network operators to ensure efficient service. By optimising their locations in real time the UAV infrastructure can achieve greater service efficiency, while also reducing overheads.
\end{enumerate}

While they introduce a variety of benefits to wireless networks the UAVs are limited by their on-board battery life, which restricts the length of time that a given UAV can stay in the air. As a consequence of this the UAV-mounted infrastructure can only provide temporary service to an area of interest, unless a solution is implemented to address the battery life issue. As they are a new technology that has not yet seen commercial adoption the issue of the limited UAV battery life is inadequately explored in the research literature. The wireless community has published a variety of works on the subject of optimising the energy efficiency of individual UAVs through the optimisation of select parameters such as trajectory or transmit power, for example \cite{Zeng_2017}. While optimising individual UAVs can improve their performance during their flight, it is not sufficient for creating a viable UAV network. 

There is currently insufficient insight into how to design a cellular network which uses UAVs to augment communications infrastructure and which accomodates the fact that UAVs are incapable of staying airborne for long periods of time. In this article we explore several approaches for a network operator to address the UAV battery lifetime issue and design a UAV network that enables continuous wireless coverage. We evaluate the sort of network performance that can be achieved by implementing these solutions, and we discuss the relative strengths and weaknesses of each approach. Furthermore, we provide a high level overview of the developments being made in the field of battery technology, and demonstrate how they may improve UAV performance in the foreseeable future.

\section{UAV Characteristics}

We begin by providing a short discussion on the type of UAVs currently available on the civilian market and which variants we expect to be used for flying infrastructure.
UAVs vary greatly in size, with the smallest UAVs weighing less than $\unit[1]{kg}$ and fitting comfortably inside personal bags, while the larger UAVs are the size of manned aircraft. The variance in size also corresponds to different regulations and restrictions. The FAA (Federal Aviation Administration) and the EASA (European Aviation Safety Agency) currently restrict UAVs with a take-off weight below $\unit[25]{kg}$ to operate at heights below $\unit[120]{m}$, which corresponds to unregulated airspace which manned aircraft do not operate in, whereas the larger UAVs are required to use regulated airspace and coordinate with air traffic control. Companies such as Google \cite{Google_2018} have expressed interest in using large, higher-altitude UAVs for providing basic wireless connectivity to remote areas with limited existing infrastructure. For dense, urban areas the small, low altitude UAVs are more appropriate, as they are safer to use due to their small size and they can fine-tune their positioning in 3D space in a manner that is unavailable to the larger aircraft.

Low altitude UAV designs can be separated into two categories based on their method of flight. The first category of UAVs is referred to as fixed-wing, and corresponds to an airplane design, where UAVs have wings which generate lift from air passing underneath. The second category is the rotor-wing, where the UAVs have several rotors with propellors which push air downwards and generate enough thrust to counter the force of gravity on the UAV. Both designs have advantages and disadvantages. The advantage of the fixed-wing design is that it allows the UAV to fly with less thrust from its motors due to the behaviour of aerodynamic flow. Less thrust needed to fly corresponds directly to less energy consumed, which means a longer flight time for a given battery. The disadvantage of this design is that the UAV must always be moving forward at a certain minimum speed to generate enough thrust to stay in the air, and as a consequence it is impossible to keep the UAV hovering above a certain location of interest. It also means that the UAV requires a large open area for take-off and landing, and in a dense urban environment open areas of suitable size may not be available. Because of this, we consider the rotor-wing UAV design to be the most appropriate for operating in an urban environment.

\begin{figure}[t!]
\centering
	\includegraphics[width=.45\textwidth]{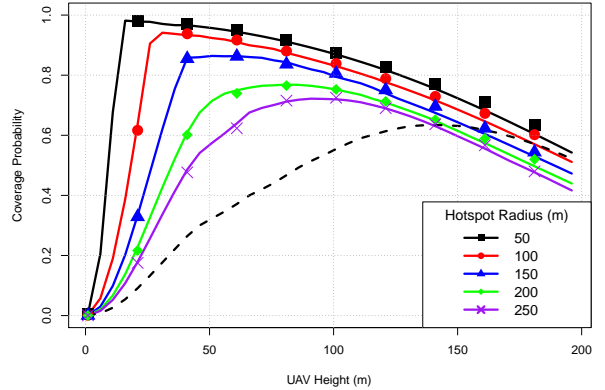}
    \vspace{-5mm}
	\caption{
	Coverage probability of a UAV network as a function of UAV height and the radius of the user hotspot area that the UAV is expected to cover. The dashed line denotes the performance as the radius tends to infinity.
	}
	\label{fig:ClusterRad}
	\vspace{-3mm}
\end{figure}

In our previous work \cite{Galkin_20182} we have explored the performance of a low altitude, rotor-wing UAV network operating above user hotspots. We demonstrated that UAVs can leverage their height to find a performance sweet-spot which balances their ability to deliver a wireless signal to a typical user while also minimising the amount of interference the user experiences. The height which gives this optimum performance is a function of the density of the UAVs, their antenna configuration, the LOS-blocking buildings in the environment, and also the size of the user hotspots themselves. These results are illustrated in \Fig{ClusterRad}. We expect that an intelligent UAV network will position UAVs at the heights which give the best performance: the consequence of this is that the UAVs will have to expend a certain amount of their total battery power on getting into position to serve the users, with the exact amount of battery power (and the resulting battery life left) being highly dependent on the environmental parameters. We discuss these issues next.

\section{UAV Battery Life Today}

\begin{table}[b!]
\vspace{-3mm}
\begin{center}
\caption{Model Parameters}
\begin{tabular}{ |c|c| } 
 \hline
 Parameter & Value \\ 
 \hline
 Docking Station Density & \unit[1]{$/\text{km}^2$} \\
 Hotspot Density & \unit[5]{$/\text{km}^2$} \\
 Optimum UAV Heights & \unit[15-500]{m} \cite{Galkin_20182}\\
 Mean Docking Station Height & \unit[30]{m} \\
 UAV Horizontal Velocity & \unit[8]{m/s} \\
 UAV Ascent Velocity & \unit[2]{m/s} \\
 UAV Descent Velocity & \unit[1.5]{m/s} \\
 UAV Horizontal Power Consumption & \unit[206.02]{W}\cite{Suman_2018}\\
 UAV Ascent Power Consumption & \unit[249.01]{W}\\
 UAV Descent Power Consumption & \unit[212.46]{W}\\
 UAV Hover Power Consumption & \unit[221.27]{W}\\
 UAV Battery Energy Density & \unit[250]{Wh/Kg}\\
 UAV Battery Weight & \unit[0.4]{Kg}\\
 \hline
\end{tabular}
 \label{tab:table}
\end{center}
\end{table}

What sort of useful flight time can a network operator  expect from a UAV using technology that exists in the commercial market today? We consider a scenario where UAV small cells with downtilted antennas \cite{Galkin_20182} are deployed in an urban environment to supplement terrestrial infrastructure in providing users with wireless service. The UAVs are stationed at dedicated docking stations distributed on rooftops around the city, where they are kept ready for rapid deployment. When a UAV is issued with the instructions to cover a demand hotspot it takes off, travels to the hotspot and hovers above it until it has just enough power left to safely return to its docking station and recharge. The amount of energy the UAV has to spend on travel between its docking station and the hotspot depends on the UAV speed and on the distance between the two locations. We assume that the locations of the docking stations and the hotspots are random with respect to one another, as the hotspots represent unpredictable spikes in user demand. These demand hotspots may arise due to events such as outdoor markets or public demonstrations, with the hotspots varying in size from covering an area a few dozen meters across to spanning several streets. We assume that the UAVs position themselves directly above the hotspot centers at the optimum height. 

\begin{figure}[t!]
\centering
	\subfloat[]{\includegraphics[width=.45\textwidth]{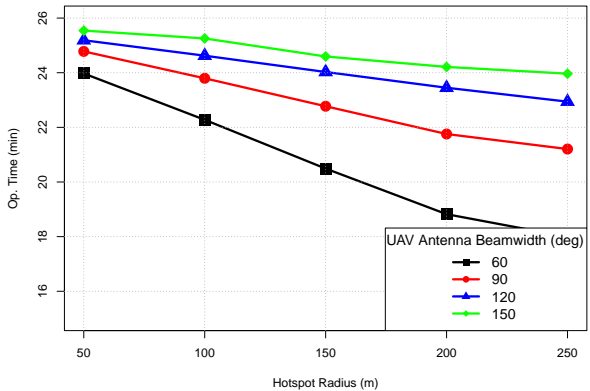}}\\

	\subfloat[]{\includegraphics[width=.45\textwidth]{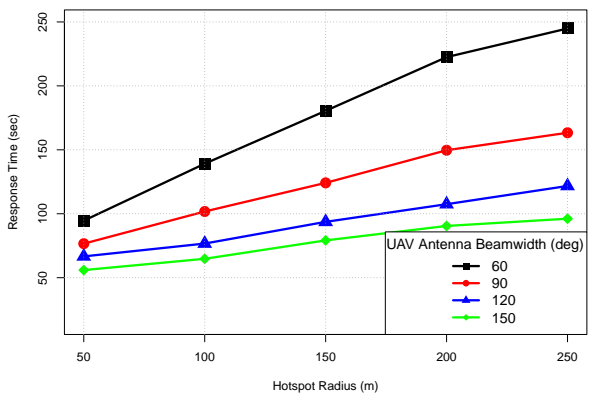}}
    
	\caption{
Operating lifetime and the response time of a UAV when it moves from its docking station to a position at the optimum height above a given hotspot, assuming horizontal velocity of $\unit[8]{m/s}$. The height is determined by the radius of the hotspot and also the beamwidth of the UAV antenna \cite{Galkin_20182}.
	}
	\label{fig:Lifetime}
	\vspace{-3mm}
\end{figure}

Unless stated otherwise, the model parameters used are given in Table \ref{tab:table}. \Fig{Lifetime} (a) shows the length of time the UAVs can hover over the hotspots before they have to return to their docking stations, as a function of hotspot radius and UAV antenna beamwidth. Larger hotspot radii and narrower antenna beamwidths correspond to higher optimum UAV altitudes, which means that the UAV must expend more battery power getting into position and must preserve more battery power for the return trip. We can see that the UAVs will have 15-25 minutes of useful flight time on a single battery before they have to recharge. Given that an outdoor event which creates user hotspots may last several hours it is clear that the operator may wish to take steps to ensure that UAV infrastructure can stay in the air for a sufficiently long period of time. Note that, while they may not be able to stay in the air for very long, the UAVs are capable of moving quite quickly through the environment, \Fig{Lifetime} (b) shows the average time it takes for a UAV to travel from its docking station to its assigned operating point. Given this rapid response time and short operating time the UAVs available today may be most suited for use in emergency scenarios where a device needs to transmit or receive critically important data quickly, but not necessarily for an extended period of time.

\section{UAV Swapping}

\begin{figure}[t!]
\centering
	\subfloat[Cycling through multiple UAVs to cover a hotspot]{\includegraphics[width=.45\textwidth]{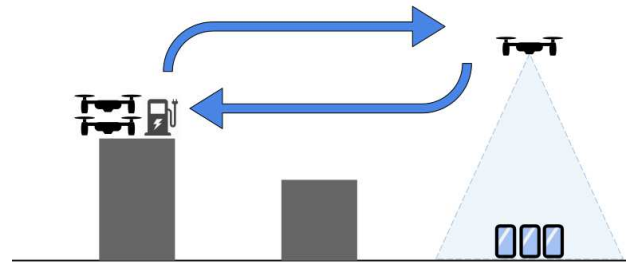}}\\
	\subfloat[Hotswapping batteries of a single UAV covering a hotspot]{\includegraphics[width=.45\textwidth]{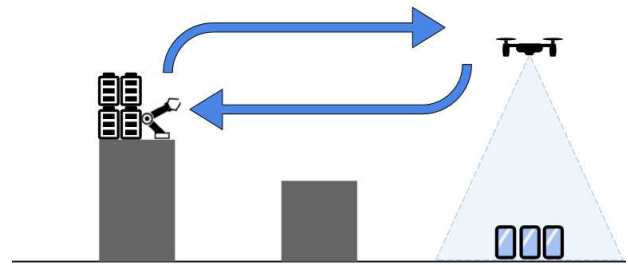}}\\
    \subfloat[Using lasers to wirelessly power a UAV]{\includegraphics[width=.45\textwidth]{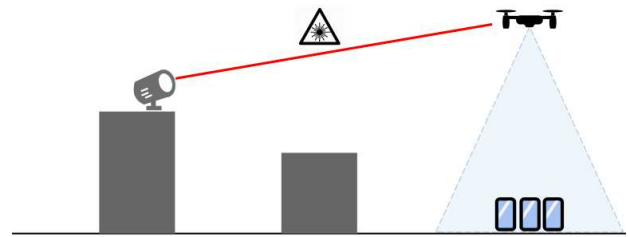}}
	\caption{
   Proposed UAV battery management solutions.
	}
	\label{fig:SystemModel}
	\vspace{-3mm}
\end{figure}

One of the most straightforward ways of building a UAV network around the limited battery life of the UAV is to sequentially switch out low-power UAVs with ones that are fully charged, as depicted in \Fig{SystemModel} (a). In this scenario, for each UAV that is operating above a user hotspot there are several other UAVs being charged at a docking station, waiting to be deployed. When the first UAV must return to its docking station to recharge it will be replaced by a second UAV, which in turn will be replaced by a third, and so on, until the first UAV is fully charged at the docking station and is ready to be deployed again. By having a sufficiently large number of backup UAVs and by timing their deployments such that one UAV hands over its hotspot seamlessly to another UAV the network can provide continuous, uninterrupted service to an area. 

The number of backup UAVs that must be kept in a state of readiness for a given hotspot will be determined by UAV "downtime", that is, the length of time the UAV will need to travel back to its docking station, recharge, and return to the hotspot. The longer the recharge time, the more UAVs are needed to substitute it before it can deploy again. According to technical specifications from leading civilian UAV manufacturer DJI, a commercial UAV has a recharge power of up to $\unit[180]{W}$ \cite{DJI_Inspire2}. The results of our simulation suggest that as few as two backup UAVs may be required for each operating UAV in the network to ensure uninterrupted coverage. New battery charging technologies to enable faster energy transfer and reduce charging time are needed to reduce the number of backup UAVs and make the UAV network more affordable. Note that the UAV horizontal velocity does not appear to have a significant impact on the number of backups: a higher velocity allows UAVs to spend less time on travel; however, it also consumes more battery power \cite{Suman_2018}.

\section{Battery Hotswapping}
The majority of high-end UAVs nowadays are designed with external battery packs that can be detached from the UAV, thus enabling fast swapping of batteries by the UAV operator. Certain high-end models even carry two external batteries, both for safety reasons and to enable battery hotswapping. Battery hotswapping is when a UAV battery is replaced without the UAV being powered off, which allows it to return to its regular operation the moment the new battery is in place. The drawback of battery hotswapping is that it currently requires a human operator to carry out the mechanical operation of detaching the depleted battery and inserting a new one into the UAV. This introduces human labour into what may otherwise be an automated network. To address this, researchers have explored the concept of automated battery swapping stations, where robotic actuators are used to switch out batteries. The authors of \cite{Lee_2015} demonstrate a working prototype of such a station, showing how a UAV can automatically land into the docking station and have its battery swapped out within 60 seconds. 

\begin{figure}[b!]
\centering
	\subfloat{\includegraphics[width=.45\textwidth]{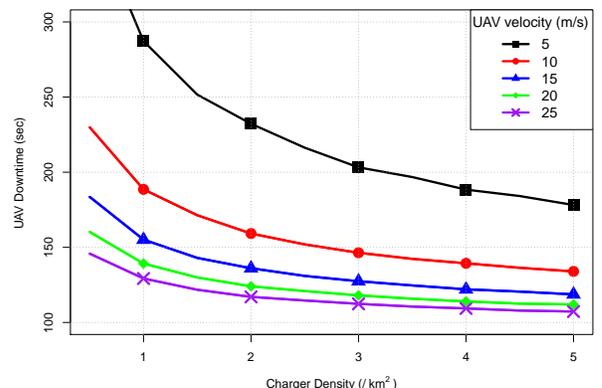}}\\
	\vspace{-6mm}
	\caption{
   Total UAV downtime under battery hotswapping, as a function of docking station density and UAV horizontal velocity.
	}
	\label{fig:Downtime}
	\vspace{-3mm}
\end{figure}

To demonstrate the benefits of this setup we consider the scenario depicted in \Fig{SystemModel} (b). A UAV provides service above a user hotspot until its battery is depleted and it must return to its docking station. There, its battery is hotswapped with a backup battery and it returns to its hotspot, while its previous battery is charged up. Instead of having several backup UAVs we have several backup batteries, which reduces the cost of the infrastructure; however, because we only have one UAV per hotspot the hotspot will not be serviced for the length of time it takes the UAV to move to its docking station, hotswap its battery, and return back. \Fig{Downtime} shows the duration of this downtime, as a function of UAV speed and density of docking stations per unit area, assuming the hotswap procedure takes 60 seconds as in \cite{Lee_2015}. We can see that the total downtime will last less than 3 minutes for the majority of the UAV velocities. The 3 minutes of downtime for every 20-25 minutes of operating time (as per \Fig{Downtime}) may be accepable if the hotspot corresponds to regular user data traffic; for emergencies or other scenarios where the data is time-critical the operator may wish to have a backup UAV available, as in the previous section. 

\section{Wireless Power Transfer}
Battery hotswapping appears to be a viable solution to the limited UAV battery life. However, it still requires UAVs to regularly move between their serving location and a docking station, which reduces the operating efficiency of the network. An alternative approach is to wirelessly transfer power to the UAVs to enable them to stay in the air without needing to recharge. A variety of techniques for wirelessly transferring power to a UAV have been researched. These can be roughly separted into two categories: electromagnetic field (EMF) charging and non-EMF charging. EMF charging refers to using electro-magnetic fields to transfer energy, using magnetic induction or similar. These techniques work across a very short range (in the order of centimeters) and are incapable of transferring sufficient energy quickly enough to compensate for the energy consumption of the airborne UAV. Non-EMF refers to using photo-voltaic (PV) cells to charge UAVs. For large UAVs these PV cells would harness solar power to keep the UAV flying; however, for smaller UAVs solar power is inappropriate due to the smaller cross-section of the PV cell. Instead, we consider the case where the PV cells have energy beamed to them using lasers. The company Lasermotive has demonstrated a working prototype of a UAV which is kept in the air for over 12 hours nonstop using a kilowatt laser which transmits a beam of energy at a specially designed PV panel on the UAV \cite{Ouyang_2018}. The difficulty with using lasers for energy transmission is that the lasers require an unobstructed LOS to the UAV to be able to reach it with their beam. In an urban environment with buildings of varying heights it may be difficult to guarantee a LOS link between a given UAV and its laser transmitter. 

\begin{figure}[t!]
\centering
	\subfloat[]{\includegraphics[width=.45\textwidth]{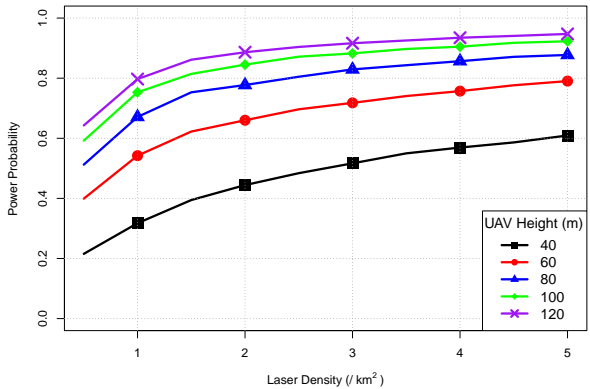}}\\
	\subfloat[]{\includegraphics[width=.45\textwidth]{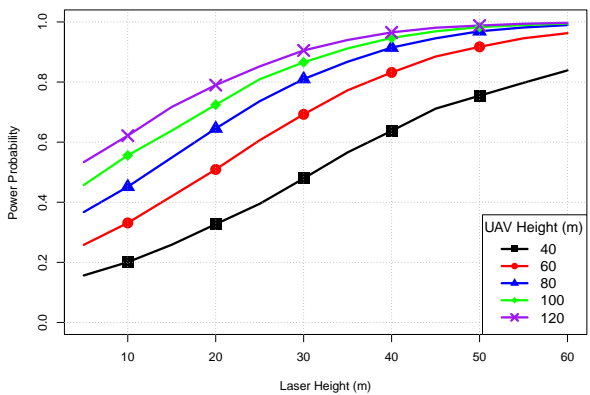}}
	\caption{
   Probability that a UAV hovering above a hotspot can be successfully charged by the nearest laser transmitter to it.
	}
	\label{fig:LaserPower}
	\vspace{-3mm}
\end{figure}

We explore the viability of radiative power transfer in the scenario depicted in \Fig{SystemModel} (c). We assume a number of laser transmitters are mounted on rooftops in a city. A UAV deployed above a hotspot will attempt to establish a LOS link to the nearest transmitter and have the transmitter beam power to it. The expression for the energy propagation of the laser beam is given in \cite{Ouyang_2018}, and the beam is assumed to be deactivated if there is a LOS obstruction for safety reasons. In \Fig{LaserPower} we give the probability that the UAV will receive sufficient power from the laser beam to negate its power consumption (and thus remain in the air indefinitely), for varying densities and heights of the laser trasmitters. As the figure shows, the probability will depend significantly on how high the UAV is above ground, with greater heights making it more likely that the UAV can be wirelessly charged. The issue is that relying on wireless charging for the UAV therefore limits the heights that the UAV can operate at. Furthermore, the current legal height limit for UAVs in Europe and the USA is approximately $\unit[120]{m}$ which, according to our results, will not allow for guaranteed wireless charging unless the laser transmitter density is very high or the transmitters are positioned high above ground. Another issue with the laser transmitter is that it can only power a single UAV at a time, as it has to mechanically steer its laser beam towards the UAV. This limits the number of laser-powered UAVs that can be deployed in an area, as each UAV has to have its own dedicated laser transmitter when operational. Using laser transmitters to wirelessly power UAVs may be a good solution for UAVs that operate at higher altitudes than those currently envisioned by aviation authorities; however, for low-altitude UAVs operating in built-up areas it may not be the most practical solution.

\section{Battery Energy Density Improvements}

\begin{figure}[b!]
\centering
	\subfloat{\includegraphics[width=.45\textwidth]{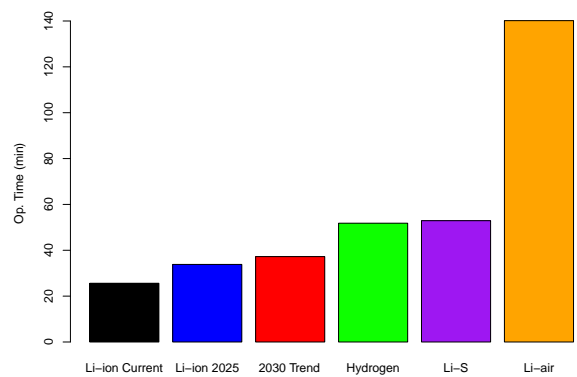}}\\
	\vspace{-6mm}
	\caption{
   Simulated UAV operating time for different battery types. Black denotes the lithium-ion technology today, blue denotes the predicted plateau of li-on batteries, red gives the predicted battery performance by 2030 following the $3\%$ rule, green denotes hydrogen cells, purple denotes lithium-sulfur batteries, and orange is lithium-air.
	}
	\label{fig:Comparison}
	\vspace{-3mm}
\end{figure}

Battery technology continues to advance at a steady pace, spurred on by the demand for greater energy density from the consumer electronics and electrical vehicle sectors. This improvement affects the cost of battery manufacture, the safety of the materials used, and the energy density of the batteries. Given that UAVs are significantly affected by their limited flight time we are particularly interested in the battery energy density, and how its improvement will improve the performance of the UAV network. 

The authors of \cite{Zu_2011} suggest that historical improvement of battery energy density can be approximated as a steady $3\%$ performance increase per year, which the authors point out is far too slow to satisfy the demands of the new, emerging technologies. Current commercially available UAVs use lithium-ion batteries with an energy density in the order of $\unit[250]{Wh/kg}$, and the research discussed in \cite{NOZAWA_2018} suggests that lithium-ion batteries may have their energy density improved by 20-30\% within the next 5 years, reaching a performance ceiling by around 2025. So-called solid state batteries which use solid electrolytes are expected to contribute to this performance growth. Sodium-ion batteries are predicted to be one of the new battery variants to act as an alternative to lithium-ion \cite{Vaalma_2018}, as the required materials are much more abundant than those used for lithium-ion batteries, which means the battery manufacturing cost would be far less vulnerable to market fluctuations. Unfortunately, sodium-ion batteries have a lower energy density than lithium-ion batteries so it is unlikely they will be a key driving technology for UAV networks. Three battery technologies on the horizon that do promise an improvement in energy density are the hydrogen fuel cell, the lithium-sulfur battery and the lithium-air battery, with a theoretical energy density of approximately $\unit[490]{Wh/kg}$ \cite{Plaza_2017}, $\unit[500]{Wh/kg}$ \cite{Service_2018} and $\unit[1,300]{Wh/kg}$ \cite{Rahman_2014}, respectively. Unfortunately, these technologies have drawbacks which delay their adoption and commercialisation. There are concerns with the safety of both hydrogen fuel cells and lithium-sulfur batteries, while lithium-air batteries are known to be very vulnerable to exposure to the outside environment. Because of these drawbacks it is difficult to make an estimate on the dates when the new batteries may be adopted into UAV networks and the real-world performance these batteries will have. In \Fig{Comparison} we aggregate the published findings to show the predicted operating time of UAVs in the coming years. A conservative estimate following the $3\%$ annual performance increase suggests that UAVs may be able to fly in the order of 40 minutes by 2030 if hydrogen fuel, lithium-sulfur or lithium-air batteries are not commercialised by then. If they are, UAVs may be able to operate in the air for 1-2 hours at a time without needing a recharge in the not-too-distant future.

\section{Conclusion}

In this article we considered several network design solutions that can be integrated into a cellular network to enable UAV infrastructure to serve user hotspots for extended periods of time. We considered the possibility of a network of UAV charging stations being deployed on rooftops in a city, and we demonstrated that as few as three UAVs are needed to provide continuous, uninterrupted coverage of a given area, when one UAV flies above a target hotspot and the other two are waiting to be deployed at their charging station. Another option we investigated is the use of battery hotswapping, where a docking station with a mechanical actuator switches out the depleted battery of a UAV with a new one. Our results suggested a total downtime of below three minutes, using existing technology. We also investigated the option of powering the UAVs using lasers. Our results suggest that laser power may be unsuitable for low-altitude UAVs in cities, due to the presence of LOS-blocking buildings. Finally, we investigated the developments being made in battery capacity. New technologies on the horizon promise to extend the UAV flight time to 1-2 hours, which should greatly alleviate the battery lifetime limitation of UAV infrastructure.

\section*{Acknowledgements}
This material is based upon works supported by the Science Foundation
Ireland under Grants No. 14/US/I3110 and 13/RC/2077.

\ifCLASSOPTIONcaptionsoff
  \newpage
\fi



\bibliographystyle{./bib/IEEEtran}
\bibliography{./bib/IEEEabrv,./bib/IEEEfull}

\end{document}